\documentstyle[11pt]{article}
\newcommand{\be}{\begin{equation}}
\newcommand{\ee}{\end{equation}}
\newcommand{\bea}{\begin{eqnarray}}
\newcommand{\eea}{\end{eqnarray} }
\def\IR{{\hbox{{\rm I}\kern-.2em\hbox{\rm R}}}}
\begin{document}
\pagestyle{plain}
\title{Topological Roots of Black Hole Entropy\thanks{To appear
in the Proceedings of the Lanczos Centenary Conference, edited
by J.D.~Brown, M.T.~Chu, D.C.~Ellison, and R.J.~Plemmons (SIAM,
Philadelphia)}}
\author{Claudio Teitelboim\\
Centro de Estudios Cientificos de Santiago\\
Casilla 16443, Santiago 9, Chile \\
and\\
Institute for Advanced Study\\
Olden Lane, Princeton, NJ 08540 USA}
\date{April 27, 1994}
\maketitle
\begin{abstract}We review the insights into black hole entropy that
arise from the formulation of gravitation theory in terms of dimensional
      continuation. The role of the horizon area and the deficit angle
      of a conical singularity at the horizon as canonically conjugate
      dynamical variables is analyzed. The path integral and the extension
      of the Wheeler-De Witt equation for black holes are discussed.
\end{abstract}

\section{Introduction} Boltzmann's formula
\be
  S = {\rm log}W
\ee
is a cornerstone of statistical mechanics. It relates $S$, the
macroscopic entropy of a system, to $W$, the number of microscopic
states of the system which have the same given macroscopic properties.

An outstanding problem in gravitation theory is to express the black hole
entropy of Bekenstein and Hawking
\be
  S = \frac{1}{4G\hbar} ({\rm horizon~area})
\ee
in terms of (1). This poses two questions, namely,

(i) What are the microscopic states?

(ii) How many are there?

\noindent It turns out that formulating gravitation theory in terms
of dimensional
continuation provides an answer to the first question
and suggests an answer to the second.

The idea is the following. One analyzes the gravitational action
keeping in mind that in two spacetime dimensions it reduces to a
topological invariant, the Euler class. This has two rewards. First,
the dimensional continuation of the Gauss--Bonnet theorem shows that
the black hole entropy itself is the dimensional continuation of the
Euler class of a small disk centered at the horizon. Second, since
the Euler class of the small disk is still well defined when one
allows for a cusp (conical singularity) within the disk, it is
natural to allow for new degrees of freedom to be admitted in the path
integral that correspond to the possibility of a conical singularity.

One then finds that the ``deficit angle" of the cusp and the horizon
area are canonically conjugate. Summing over all horizon areas yields
the black hole entropy. This provides an answer to question (i) above.
However, a ``microscopic explanation" for the exponential weight in the
integration measure for the surface degrees of freedom, or equivalently
for the $\hbar^{-1}$ dependence in (2), is still lacking.

Thus the answer to the second question is not provided but only suggested
by the present analysis: It would seem natural to attempt to obtain the
black hole entropy as the ``number of states within a very small
two--dimensional disk". This has not been done at the moment of this
writing.

The plan of this report is the following. Sections 2, 3, and 4 review the
treatement of the action and the entropy in terms of
dimensional continuation. These sections are based on joint work with
M.~Ba\~nados and J.~Zanelli and follow closely Ref.~\cite{BTZ}. Section
5 discusses the relationship of the surface degrees of freedom with the
propagator whose trace is the partition function. This section is based on
joint work with S.~Carlip and follows Ref.~\cite{CT}.

\section{The Action as the Dimensional Continuation of the Euler Class}
If one considers a two dimensional manifold $M$ with
boundary $\partial M$, the Gauss--Bonnet theorem reads
\begin{equation}
\frac{1}{2}\int_M \sqrt{g} g^{\mu \nu} R^{\alpha}_{\;\; \mu
\alpha \nu}d^2x - \int_{\partial M} \sqrt{g}K d^1x =
2\pi\chi(M) \ .
\label{1}
\end{equation}
The integer $\chi(M)$ on the right hand side of (\ref{1}) is the
Euler number of $M$ and depends solely on its topology. One has
$\chi=1$ for a disk and $\chi =0$ for an annulus. We will refer
to the sum of integrals appearing on the left side of (\ref{1})
as the Euler class of $M$. The Gauss--Bonnet theorem then says
that the Euler class of $M$ is equal to $2\pi$ times its Euler
number.

If one varies the integral over $M$ in (\ref{1}) one finds, by
virtue of the Bianchi identity, that the piece coming from the
variation of the Riemann tensor yields a surface term. This
surface term exactly cancels the variation of the surface
integral appearing in the Euler class. On the other hand,
because of the special algebraic properties of the Riemann
tensor in two spacetime dimensions, the contribution of the
variation of $\sqrt{g}g^{\mu \nu}$ is identically zero.  This
is a poor man's way to put into evidence that the Euler class is
``a topological invariant", the real work is to show that the
actual value of the sum of integrals is $2\pi \chi$.

Now, the Hilbert action for the gravitational field in $d$
Euclidean spacetime dimensions may be written as
\begin{equation}
I_H = \frac{1}{2}\int_M \sqrt{g} g^{\mu \nu}
R^{\alpha}_{\;\;\mu \alpha \nu}d^dx - \int_{\partial M} \sqrt{g}
K d^{d-1}x \ .
\label{2}
\end{equation}
[One integrates $\exp(+I)$ in the Euclidean path integral.
We have set $8\pi G = 1$.]
This action has the same form as the Euler class of two
dimensions, with the change that now the integrals, and the
geometric expressions appearing in them, refer to a spacetime of
dimension $d>2$. For this reason, one says that the Hilbert
action is the dimensional continuation of the Euler class of two
dimensions. After dimensional continuation, the Euler class
ceases to be a topological invariant. While it is still true
that the variation of the Riemann tensor in (\ref{2}) yields a
surface term, this surface term no longer cancels the variation
of the integral of the extrinsic curvature. Rather, the sum of
the two variations vanishes only when the intrinsic geometry of
the boundary is held fixed. Moreover, the contribution to the
variation coming from $\sqrt{g}g^{\mu \nu}$ gives the
Einstein tensor, which is no longer identically zero, and hence
the demand that it vanishes is not empty but gives the Einstein
equations.

This reasoning applies also to the natural generalization of the
Hilbert action to higher spacetime dimensions, the
Lovelock action \cite{Lovelock}. This action, which keeps the
field equations for the metric of second order and hence does
not change the degrees of freedom, can also be
understood in terms of dimensional continuation
\cite{Zumino,Teitelboim-Zanelli}. For a spacetime of dimension
$d$, the generalized action contains the dimensionally continued
Euler classes of all even dimensions $2p<d$. Thus, the Hilbert
action with a cosmological
constant may be thought of as coming from dimensions $2p=2$ and
$2p=0$, respectively.

The analog of the Hilbert action given by (\ref{2}) is
\begin{equation}
I_{L} = \sum_{2p<d} \frac{\alpha_p}{2^{2p} \, p! } (I_{L}^p +
B^p) \ ,
\label{10'}
\end{equation}
with
\begin{equation}
I_{L}^p=  \int_M\sqrt{g} \delta_{[\alpha_1 \cdots
\alpha_{2p}]}^{[\beta_1 \cdots \beta_{2p}]} R^{\alpha_1
\alpha_2}_{\beta_1 \beta_2} \cdots R^{\alpha_{2p-1}
\alpha_{2p}}_{\beta_{2p-1} \beta_{2p}}d^dx \ .
\label{11}
\end{equation}
(Here the totally antisymmetrized Kronecker symbol is normalized
so that it takes the values 0, $\pm 1$.)

The boundary term $B^p$ is the generalization of the integrated
trace of the extrinsic curvature in (\ref{2}). It is given by
\begin{equation}
B^p = \frac{-2}{d-2p} \int_{\partial M} d^{d-1}x
g_{ij}\pi^{ij}_{(p)} \ .
\label{12}
\end{equation}
Here $\pi^{ij}_{(p)}$ is the contribution of (\ref{11}) to the
momentum canonically conjugate to the metric $g_{ij}$ of
$\partial M$. It may be expressed as a function of the intrinsic
and extrinsic curvatures of the boundary
\cite{Teitelboim-Zanelli}.

\section{Covariant Action versus Canonical Action. Entropy as
Dimensional Continuation}
There is another action, which differs from the $I_H$ by
boundary terms. It is the canonical action
\begin{equation}
I_C = \int ( \pi^{ij} \dot{g}_{ij} - N {\cal H} -
N^{i}{\cal H}_i) \ .
\label{3}
\end{equation}
When one studies black holes $I_C$ has a significant advantage
over the Hilbert action. It vanishes on the black hole due to
the constraint equations ${\cal H} =0= {\cal H}_i$ and the time
independence of the spatial metric.  The black hole entropy and
its relation with the Gauss-Bonnet theorem will arise through
the difference between the Hilbert and the canonical actions.

In the Euclidean formalism for black holes, it is useful to
introduce a polar system of coordinates in the $\IR^2$ factor
of $\IR^2 \times {\cal S}^{d-2}$. The reason is that the
black hole will have a Killing vector
field---the Killing time---whose orbits are circles centered
at the horizon. But, it should
be stressed that the discussion that follows is valid for a
system of polar coordinates centered anywhere in $\IR^2$. Indeed
the Killing vector exists only on the extremum and not for a
generic spacetime admitted in the action principle.

Take now a polar angle in $\IR^2$ as the time variable in a
Hamiltonian analysis. An initial surface of time $t_1$ and a
final surface of time $t_2$ will meet at the origin, which is a
fixed point of the time vector field. There is nothing wrong
with the two surfaces intersecting. The Hamiltonian formalism
can handle that. Next, divide $\IR^2$ into a small disk
$D_{\epsilon}$ of radius $\epsilon$ around the origin, and an
annulus of inner radius $\epsilon$ and outer radius that will
tend to infinity. Analysis of the boundary terms---which will
not be given here---shows that, in the limit $\epsilon\to 0$,
the Hilbert action for the annulus and the canonical action differ
only by a local surface integral at $r=\infty$. Thus we have
\be
I_H = \lim_{\epsilon \rightarrow 0} I_H[D_{\epsilon}\times
{\cal S}^{d-2}] + I_C + B_{\infty} \ .
\label{6}
\ee
Here $I_C$ is the canonical action (\ref{3}) for the annulus in the
limit $\epsilon\to 0$.

The boundary term $B_{\infty}$, which need not be explicitly
written, appears because of the different boundary conditions
at infinity for $I_H$ and $I_C$. Indeed, as stated above, the
Hilbert action (\ref{2}) needs the intrinsic geometry of the
boundary at $r=\infty$ to be fixed. On the
other hand, for the Hamiltonian action (\ref{3}) one must fix at
infinity the mass $M$ and angular momentum $J$---with a precise
rate of fall off for the fields (see, for example
\cite{Regge-Teitelboim}). If instead of $M$ one fixes its
conjugate, the asymptotic Killing time difference
$\beta$, while still keeping $J$ fixed, one must
substract $\beta M$ from (\ref{3}).

The contribution at the origin in (\ref{6}) appears precisely
because there is no boundary there in the topological sense. Indeed,
the canonical action introduces an additional structure, the
time vector field which has a fixed point at the origin. This makes
it not covariant. The boundary term is brought in in order to
restore covariance.

Thus, if we drop
$B_{\infty}$, we obtain the improved covariant action,
\begin{equation}
I = \lim_{\epsilon \rightarrow 0} I_H[D_{\epsilon} \times
{\cal S}^{d-2}] + I_C \ ,
\label{7}
\end{equation}
which is suited for fixing $M$ and $J$ at infinity. The
action (\ref{7}) differs from expression (\ref{6}) only by a
local surface term at infinity due to the different boundary
condition there, and it is therefore as covariant as (\ref{2}).
Furthermore, (\ref{7}) is finite on the black hole and thus it
is ``already regularized''. [The Hilbert action (\ref{2}) is
infinite on the black hole because $B_{\infty}$ diverges.]

A short analysis reveals that the first term in (\ref{7})
factorizes into the product of the Euler class (\ref{1}) for
$D_{\epsilon}$ and the area of the ${\cal S}^{d-2}$ at
the origin. Thus one finds
\begin{equation}
\lim_{\epsilon \rightarrow 0} I_H [D_{\epsilon} \times
{\cal S}^{d-2}] = 2\pi \times (\mbox{area of }{\cal S}^{d-2}
)_{origin} \ .
\label{8}
\end{equation}

Consider now the value of the action on the extremum. Then it is
convenient to take the polar angle to be the Killing time,
for---in that case---the spatial geometry $g_{ij}$ is time
independent.  Furthermore, since the Hamiltonian contraints
${\cal H}={\cal H}_i=0$ hold on the extremum, the value of the
improved action (\ref{7}) for the black hole is just the
contribution of the disk at the horizon,
\begin{equation}
S =   2\pi \times (\mbox{area of ${\cal S}^{d-2}$})_{horizon} \ .
\label{10}
\end{equation}
This is the standard expression for the black hole entropy in
Einstein's theory, in the semiclassical approximation (``tree
level"). This should be the case since in (\ref{7}) $M$ and $J$
are fixed, which corresponds to the microcanonical ensemble.

Note that the overall factor in front of
the area, usually quoted as one fourth in units where Newton's
constant is unity, is really the Euler class of the two-dimensional
disk.

\section{Deficit Angle as Off--Shell Degree of Freedom. Partition
Function} On account of the Gauss--Bonnet theorem the value
of the Euler class for a disk is equal to $2\pi$ even if there is
a conical singularity (curvature localized at a point). It is
therefore natural to allow for that possibility. If there is
a ``cusp of deficit angle
$\alpha$'' at the origin of $\IR^2$, the value
of the two--dimensional integral in the Euler class (\ref{1}) is
equal to $\alpha$, whereas the line integral over the boundary
has the value $2\pi - \alpha$. The full action (\ref{7}) depends
on $\alpha$. This is most directly seen by recalling that---as
stated in (\ref{6})---the action (\ref{7}) differs from the
Hilbert action (\ref{2}) by a local boundary term at infinity. As a
consequence, if the geometry of the ${\cal S}^{d-2}$ at the cusp is
varied, one finds that the action changes by
\begin{equation}
\delta I = \alpha \delta(\mbox{area of ${\cal S}^{d-2}$ at cusp})
\ .\label{9}
\end{equation}
Equation (\ref{9}) shows that the deficit angle, which is a
property of the intrinsic Riemannian geometry of $\IR^2$, is
canonically conjugate to the area of the ${\cal S}^{d-2}$
attached to that point---an extrinsic property.

Observe that one could incorrectly believe, due to (\ref{8}), that
the action (\ref{7}) (and hence its variation) is independent of
the deficit angle $\alpha$. What happens is that there is a
boundary term in the variation of the canonical action, coming
from space derivatives in ${\cal H}$, which cancels the
variation of the surface term in the Euler class
leaving (\ref{9}) as the net change \cite{Brown-York}.

As shown by (\ref{8}), the actions (\ref{3}) and (\ref{7})
differ by a contact transformation which depends only on the intrinsic
geometry of the ${\cal S}^{d-2}$ at the origin. Therefore, if
that geometry were held fixed, both actions
correctly yield Einstein's equations and, on this basis, they
would be equally good. However, in the calculation of the partition
function (see below) one must integrate over all ``closed Euclidean
histories" keeping fixed only the data at infinity. This means that
in the semiclassical approximation one must extremize with respect
to the geometry of the ${\cal S}^{d-2}$ at the origin, instead of
keeping it fixed.

For that problem, the improved action (10) and the canonical
action (8) are not equivalent. The black hole will be an
extremum for the covariant action
(\ref{7}), because the demand that the variation (\ref{9})
vanishes yields $\alpha=0$ at all points, which is the condition
for the manifold to be metrically smooth.  This is a property
that the Euclidean black hole indeed posseses, since the empty
space Einstein equations are obeyed everywhere.  On the other
hand, the demand that the canonical action should have an
extremum with respect to variations of the area of the ${\cal
S}^{d-2}$, would yield $\alpha = 2\pi$ at the origin, which
would introduce a sort of source at the origin.

Thus, adding the Hilbert action for a small disk around the
origin to the canonical action restores covariance without
introducing sources. This addition ensures that the fixed point
can be located anywhere. This must be so since the manifold has
only one boundary, that at infinity.
In this sense the presence of a non--vanishing black hole entropy
given by (12) is a consequence of general covariance.

The preceding analysis goes through step by step for the
Lovelock theory \cite{Lovelock}.
For Euclidean black holes in $d$--spacetime dimensions, again with
topology $\IR^2 \times {\cal S}^{d-2}$ \cite{BTZ2}, the action
(\ref{7}) now reads
\be
I = \lim_{\epsilon \rightarrow 0} I_{L} [D_{\epsilon} \times
{\cal S}^{d-2}]  + I_C \ ,
\label{13}
\ee
and the entropy becomes
\begin{equation}
S = \lim_{\epsilon \rightarrow 0} I_{L} [D_{\epsilon} \times
{\cal S}^{d-2}] \ .
\label{13'}
\end{equation}
The limit (\ref{13'}) factorizes into the Euler class of the
disk, equal to $2\pi$, and a sum of dimensional continuations to
${\cal S}^{d-2}$ of the Euler classes of all even dimensions below $d-2$,
\begin{equation}
S = 2\pi \times \sum_{2p<d} \frac{\alpha_p}{2^{2(p-1)}[2(p-1)]!} S^{p-1}
\label{14}
\end{equation}
with
\begin{equation}
S^p = \int \sqrt{g}\delta_{[\alpha_1 \cdots \alpha_{2p}]}^{[\beta_1
\cdots \beta_{2p}]} R^{\alpha_1 \alpha_2}_{\;\;\;\beta_1
\beta_2}\cdots R^{\alpha_{2p-1} \alpha_{2p}}_{\;\;\;
\beta_{2p-1} \beta_{2p}} d^{d-2}x \ ,
\label{14'}
\end{equation}
where the integral is taken over the $(d-2)$--sphere at the horizon.

The Hilbert action corresponds to $2p=2$ and the corresponding
entropy is $2\pi$ times the area. The cosmological constant term
corresponds to $2p=0$ and gives no contribution to the entropy.
Expression (\ref{14}) was first given in
\cite{Jacobson-Myers}.

\section{Partition Function as Trace of Propagator over Horizon
Degrees of Freedom} In the previous sections the attention was
focused on the complete black hole spacetime. To identify more
precisely the horizon degrees of freedom it is necessary to analyze
the dynamics for a wedge between $t_1$ and $t_2$.

The first observation is that the action for the wedge will again
be given by Eq.~(10). This is just because we want to obtain the
partition function as a trace of the propagation amplitude. Equation
(11) then shows that the integration measure over the horizon
geometries has a contribution of classical order that appears
as the difference between the Hilbert and the canonical action for
a disk of vanishing radius.

It should be noted that the Hilbert action for the wedge between
$t_1$ and $t_2$ is given by $I_H({\rm wedge}) =
I_C + \pi({\mbox{area of ${\cal S}^{d-2}$ at horizon}}) + B_\infty
+ \pi({\mbox{area of ${\cal S}^{d-2}$ at infinity}})$. It differs from
(10) and is not the correct action for the wedge. This means that, after
dimensional continuation, the Hilbert action for a ``full turn
wedge" is not the same as that for a disk. Before dimensional
continuation the ${\cal S}^{d-2}$ factors are absent and the
action is the same for both configurations.

The next step is to give the boundary conditons which characterize a
wedge of an ``off--shell" black hole. At infinity they will be the
usual conditions expressing a localized distribution of matter
(see, for example, \cite{Regge-Teitelboim}). At the origin, although it
is unnecessarily complicated, we will conform to standard
practice and use Schwarzschild coordinates near $r_+$.  That is,
we write the generic Euclidean metric as
\be
ds^2 = N^2(r)dt^2 + N^{-2}(r)dr^2 + \gamma_{mn}(r,x^p)dx^mdx^n
\label{a9}
\ee
up to terms of order $O(r-r_+)$, with
\be
(t_2-t_1)N^2 = 2\Theta(r-r_+) + O(r-r_+)^2 \ .
\label{a10}
\ee
Here the $x^m$ are coordinates on the two--sphere $S^2$.  The parameter
$\Theta$ is the total proper angle (proper length divided by proper
radius) of an arc of very small radius and coordinate angular opening
$t_2-t_1$.  For this reason it will be called the ``opening
angle.''  If one identifies the surfaces $t=t_1$ and $t=t_2$,
thus considering a disk in $\IR^2$, then the deficit angle $2\pi-\Theta$
is the strength of a conical singularity in $\IR^2$ at $r_+$.  For the
moment, we assume for simplicity that $\Theta$ is independent of $x^m$;
we shall see below that this restriction may be lifted without changing
the conclusions.
It is important to emphasize that no a priori relation
between $\Theta$ and the asymptotic geometry is assumed.

Besides $\Theta$ and $N(\infty)$ we fix, as usual, the spatial geometries
${\cal G}_1$ and ${\cal G}_2$ at $t_1$ and $t_2$. The transition amplitude
depends on what is fixed in the action principle, that is, it takes the
form
\be
  K[{\cal G}_2,{\cal G}_1;\Theta;\beta]
\ee
with
\be
  \beta = N(\infty) (t_2-t_1) \ .
\ee
The asymptotic Killing time separation $\beta$
is conjugate to the total mass whereas the opening angle is conjugate to
the horizon area $A$. (Equation (13) remains valid for the wedge with
$\alpha= 2\pi - \Theta$.)

The propagator (20), regarded as a functional of ${\cal G}_2$, obeys the
differential equations
\bea
  \hbar \frac{\partial K}{\partial T} + M K &=& 0 \ ,\\
  \hbar \frac{\partial K}{\partial\Theta} - A K &=& 0 \ ,
\eea
in addition to the Hamiltonian constraints
\be
  {\cal H} K = {\cal H}_i K = 0 \ .
\ee
The amplitude
\be
  K[{\cal G}_2,M_2,A_2;{\cal G}_1,M_1,A_1]
\ee
to propagate from (${\cal G}_1$, $M_1$, $A_1$) to
(${\cal G}_2$, $M_2$, $A_2$) is related to the Laplace transform
of (20) in $\Theta$ and $\beta$,
\be
  K[{\cal G}_2,{\cal G}_1;M,A] \ ,
\ee
by
\bea
  & &K[{\cal G}_2,M_2,A_2;{\cal G}_1,M_1,A_1] \nonumber\\
  & &\qquad\qquad\qquad = \delta(M_2-M_1)
     \delta(A_2-A_1) K[{\cal G}_2,{\cal G}_1;M_2,A_2]  \ .\qquad\quad
\eea
The (microcanonical) partition function is obtained by integrating
(26) over ${\cal G} = {\cal G}_1 = {\cal G}_2$ {\em and}
$A$ for fixed $M$. In the semiclassical approximation the integral over
${\cal G}$ gives unity because the canonical action
is zero on--shell, whereas the integral over $A$ yields
\be
  Z = e^{\cal S}
\ee
with ${\cal S}$ given by (2). If one allows for a dependence of $\Theta$
on the coordinates $x^m$ of
the two--sphere at $r_+$ then $\Theta(x)$ becomes canonically conjugate
to the local area element $\gamma^{1/2}(x)$ on the two--sphere. Summing
over all $\gamma^{1/2}(x)$ gives back (28). Thus the entropy associated
with a small disk in $\IR^2$ at a given horizon location $x^m$ coincides
with the entropy per unit of area obtained from (2).

The above analysis shows that one may regard the black hole entropy as
arising from summing over all horizon geometries.  We still lack
a ``microscopic'' explanation for the exponential weight in the integration
measure for the surface degrees of freedom, or equivalently for the
$\hbar^{-1}$ dependence in (2).

Note, however, that the factor
multiplying the area in (12) comes from the action of a small disk
in $\IR^2$.  This would suggest that the entropy per unit area may arise
from counting the two--dimensional geometries within the small disk. This
would be satisfactory from the point of view of dimensional continuation:
the theory would be sending us back to its two--dimensional roots.

\section{Acknowledgements} The author would like to express his
appreciation to his coworkers Maximo Ba\~nados, Steven Carlip and
Jorge Zanelli, and to David Brown for many discussions and also for
his kind help in preparing this account. This work was partially
supported by grants 0862/91 and 193.1910/93 from FONDECYT (Chile),
by institutional support to the Centro de Estudios Cientificos
de Santiago provided by SAREC (Sweden) and a group of Chilean
private companies (COPEC, CMPC, ENERSIS, CGEI). This research was
also sponsored by CAP, IBM and XEROX de Chile.



\begin{thebibliography}{99}
\bibitem{BTZ} M. Ba\~nados, C. Teitelboim, and J. Zanelli, {\em Phys. Rev.
   Lett.} {\bf 72} (1994) 957.
\bibitem{CT} S. Carlip and C. Teitelboim (``The Off--Shell Black Hole",
   preprint November 1993, to be published).
\bibitem{Lovelock} D. Lovelock, {\em J. Math. Phys.} {\bf
   12} (1971) 498.
\bibitem{Zumino} B. Zumino, {\em Phys. Rep.} {\bf 137} (1986) 108.
\bibitem{Teitelboim-Zanelli} C. Teitelboim and J. Zanelli, {\em Class.
   Quantum Grav.} {\bf 4} (1987) L125; and in {\em Constraint Theory
   and Relativistic Dynamics}, edited by G. Longhi and L.  Lussana
   (World Scientific, Singapore, 1987).
\bibitem{Regge-Teitelboim} T. Regge and C. Teitelboim, {\em
  Ann.Phys.} (N.Y.) {\bf 88} (1974) 286.
\bibitem{Brown-York} See J.D. Brown and J.W. York, {\em Phys. Rev.}
  {\bf D47} (1993) 1420, for a lucid discussion of this point.
  These authors assume $\alpha =0$ from the outset and do not
  obtain it as an equation of motion. They do not discuss the role
  of the Gauss-Bonnet theorem.
\bibitem{BTZ2} M. Ba\~nados, C. Teitelboim and J.Zanelli, {\em
  Phys. Rev.} {\bf D49} (1994) 975.
  \bibitem{Jacobson-Myers} T. Jacobson and R. Myers, {\em Phys.
  Rev. Lett.} {\bf 70} (1993) 3684.

\end{thebibliography}
\end{document}